\documentclass[journal]{IEEEtran}
\usepackage{graphicx,amsmath,amsfonts,amssymb,amsthm,geometry,afterpage}
\usepackage{float,dblfloatfix}
\usepackage{authblk}  
\usepackage{booktabs,tabularx,adjustbox,rotating}
\usepackage{algorithm}
\usepackage{algpseudocode}
\usepackage{longtable}
\usepackage{xcolor}
\title{A New Three-Players Auction Bridge with Dynamic Opponents and Team Members}

\author{Sourish Sarkar}
\author{Aritrabha Majumdar}
\author{Moutushi Chatterjee}
\affil{Indian Statistical Institute, Bangalore\\
\texttt{sourish.sarkar13@gmail.com}, \texttt{bmat2311@isibang.ac.in}, \texttt{tushi.stats@gmail.com}}

\date{}

\begin{document}
\maketitle

\begin{abstract}
   This article presents a new three-player version of the bridge playing card game for the purpose of ending fixed partnerships so that the play can be more dynamic and flexible. By dynamically redefining team makeup in real time, this game design increases unpredictability and forces players to repeatedly update strategy. A novel scoring system is introduced to reduce biases present in conventional rule-based games by favoring fairness via reward systems that enforce tactical decision making and risk assessment. Being subject to regular bridge rules, this version tests players to collaborate without fixed friendships, requiring fluid adjustment and adaptive bidding behavior in real time. Strategic issues involve aggressive and defensive bidding, adaptable playing styles, and loss-seeking strategies specific to the three-player structure. The article discusses probabilistic issues of bidding, trump and no-trump declarative effects, and algorithmic methods to trick-taking. Simulation outcomes illustrate the efficiency of diverse strategies. The game's architecture is ideal for competitions and possibly influential in broadening entry pools for tournament card games.

\end{abstract}


\section{Introduction}
{  Amongst all the so-called multi-player games, the Bridge, particularly Auction Bridge game, has grabbed the attention of players as well as researchers over the years.  DeLooze and Downey~\cite{inproceedings} have rightly observed that Bridge and other multi-player games are specially challenging and hence are pursued by many game theory enthusiasts. Interestingly, Bridge happen to be one of the very few zero-sum games, for which the expert human players have so-far outperformed the artificial intelligence (refer Yeh et al.~\cite{8438937}). Ando et al.~\cite{10.1007/3-540-45579-5_23} proposed a hypothetical reasoning mechanism for interaction among agents in the context of auction in the contract bridge game to maximize gain by cooperating with the partner while competing with the opponent minimizes the loss. Wu et al.~\cite{5351676} observed that spectrum auction differs from regular auctions due to the very fact that unlike usual auctions, which are quantity limited, spectrum auctions are interference limited. They used this information to develop a multi-winner spectrum auction game which can be further used as a feature in wireless communication. 

This particular variation will change how bridge is played at the \textbf{international level}. In a usual auction bridge tournament, only \textbf{two countries} compete against each other. But in this variant, \textbf{three countries} will compete, significantly reducing the number of games, along with more subtleties and in-game dynamic relationships.

Bidding Strategies hold a key role in a Bridge game and hence have been explored extensively in  literature. Often a successful bidding requires cooperative decision based on partial information. Yeh et al~\cite{8438937} have proposed a Deep Reinforcement Learning based Bridge bidding. However, since this model requires a non-competitive assumption that the opponent always bids PASS, which may not be a plausible assumption in most of the practical scenarios,  Chen et al.~\cite{10960098} have used the concept of  Long Short-Term Memory (LSTM)  of time series models. 
}

In the present article, we propose  a variant of the usual four-auction bridge, {   which can accommodate  } dynamic opponents and dynamic team members. {  Thus,} unlike traditional bridge (refer Whitehead~\cite{whitehead1930auction}), where players compete in fixed partnerships, this version introduces variability in team composition and opponents, ensuring that players cannot rely solely on long-term coordination or memorized conventions with a fixed partner. This dynamic element enhances unpredictability and challenges players to adapt their strategies based on the changing landscape of the game.

This article introduces a novel three-player variant of auction bridge that removes fixed partnerships, replacing them with dynamically changing teammates and opponents. This innovation adds a fresh strategic layer to the classic game, demanding greater adaptability and real-time decision-making. A major highlight of the variant is its restructured point system, carefully designed to remove biases present in traditional bridge scoring. Whereas conventional systems may favor long-standing partnerships or specific playing styles, the new scoring approach promotes fairness by rewarding players for skill, flexibility, and well-calculated risks.

In this version, the standard rules of bridge largely remain intact during gameplay, but the dynamics shift significantly due to the evolving team compositions. Players can no longer rely on pre-established conventions or familiarity with a fixed partner. Instead, they must quickly assess their temporary partner’s playing style, make optimal decisions with incomplete information, and be capable of both declarer and defensive play. The ability to adjust strategies fluidly becomes central to success, pushing players toward a more holistic mastery of the game.

One of the most innovative features of this variant is its bidding phase. In traditional bridge, the bidding process is a structured communication tool between fixed partners. However, in this dynamic format, the uncertainty of temporary partnerships makes communication less predictable and riskier. Players must evaluate how much information to reveal, balance the potential benefits of aggressive bidding against the risks of failing their contract, and anticipate their current partner’s level of play. The modified scoring system enhances this dynamic by rewarding calculated risk-taking, making the bidding phase both more challenging and rewarding.

This variant also promotes fairness by reducing the advantages enjoyed by players with well-rehearsed partnerships. By rotating teams and eliminating pre-set coordination, the game emphasizes individual skill, adaptability, and quick strategic thinking. Every match becomes a test of a player's ability to understand new teammates, anticipate their decisions, and adjust gameplay accordingly.

Strategically, players are encouraged to think probabilistically and assess risks with care. Since the team compositions vary continuously, each round offers a unique strategic landscape. Unlike standard bridge, where players may rely on pre-established conventions, this format simulates real-world decision-making, where partners may be unknown or change frequently, and information is incomplete. Thus, success requires strong situational awareness and the ability to thrive under uncertainty.

This three-player bridge variant revitalizes a traditional game by integrating dynamic team structures and a fairer scoring system. It retains the core mechanics of bridge while introducing elements that make the game more inclusive, competitive, and mentally stimulating. Whether for seasoned bridge players seeking a new challenge or newcomers aiming to sharpen their decision-making skills, this variant offers a rich, thought-provoking experience with endless variability and depth.

\section{Rules of the Proposed Game}
Each player in this {  proposed} variant of auction bridge must either place a bid or choose to pass, with the exception that the first player to bid is not permitted to pass. This ensures that the game begins with an active bid, preventing immediate stagnation. The bidding follows a strict ascending hierarchy of suits, which is ordered as follows: clubs, diamonds, hearts, spades, and finally, no-trump. The minimum starting bid is 1, and the highest possible bid a player can make is 7.

If a player bids a suit, that suit is considered the trump suit for the round, meaning that cards from that suit have a higher ranking than other suits in the trick-taking phase. The trump cards hold special significance in gameplay, allowing strategic advantages to players who secure the right to name the trump suit. The ranking of cards follows the traditional order, with numerical cards followed by the face cards in increasing value—10, Jack (J), Queen (Q), King (K), and Ace (A) being the highest. If a player bids "no-trump," then no suit is designated as the trump, making the gameplay more reliant on pure skill rather than the advantage of a trump suit.

\section{Scoring Scheme as per Auction Bridge}
The scoring system of the proposed game follows the algorithm given below:

\noindent \textbf{   Until the 6\textsuperscript{th} \emph{trick} is dropped}, no point is registered. After that,
\\
\textbf{If no \emph{double} is given:} The player scores on the basis of the \emph{bid} the player has made. But if the player is unable to secure the \emph{bid} he/she has made, then each opponent scores 25 points per overtrick.
\\
\textbf{If \emph{double} is given:} The player scores based on the \emph{bid} the player has made, and the score is doubled. Moreover, the \emph{bidder} receives 25 points per \emph{overtrick}. Also, the \emph{bidder} gets 25 points \emph{insult bonus}. On the other hand, each opponent gets 50 points per undertrick.
\\
\textbf{If \emph{redouble} is given:} Everything mentioned in \emph{\textbf{double}} gets doubled.
\\
\textbf{Slam Bonus:} If a player/team scores 12 tricks, a 50-point bonus is offered. If the player/ team manages to grab all 13 tricks, a bonus of 100 points is given. The scaling is made according to \emph{\textbf{double}} or \emph{\textbf{redouble}}.
\\
Clubs, diamonds, hearts, spades, no trump carries 3, 3.5, 4, 4.5, 5 points for each \emph{trick}.\\ 
In A game of bridge, both the partners have equal contribution to secure the score. That's why we have halved the score for each \emph{individual player}.
\subsection{Honors Scoring}

Honors are scored as follows:

\begin{itemize}
    \item \textbf{Four trump honors in one hand}: 80 points  
    \begin{itemize}
        \item (A, K, Q, J, 10) -- Any 4 for a specific trump suit.
    \end{itemize}
    
    \item \textbf{Five trump honors (or four aces in No Trump) in one hand}: 100 points  
    \begin{itemize}
        \item (A, K, Q, J, 10) -- All 5 for a specific trump suit.
        \item (Four Aces) -- If playing No Trump.
    \end{itemize}

    \item \textbf{Additional honor in partner's hand, or for three plus between both hands}: 10 points each  
        \textbf{Case 1:} Any one or more honor trump cards in the partner’s hand (A, K, Q, J, 10) receive 10 points each.

        \textbf{Case 2:} If the bidder's hand and partner’s hand together contain 3 or more honor trump cards, they receive 10 points each.
\end{itemize}

Note that any \textbf{\emph{under the line}}, \textbf{\emph{over the line}}, \emph{\textbf{rubber bonus}}, is NOT given.\\
The \emph{\textbf{under the line}} and \emph{\textbf{above the line}} points are solely based on the usual \textbf{E-W} and N-S team setup. But in the 3-player setup, the player with the highest bid gets the last hand. So, the teams are variable throughout the game. To explain this in even greater detail, we can consider the least possible call a player can make; i.e 1 \emph{clubs}. In practice, the probability of calling 1 \emph{clubs}, 1 \emph{diamonds}, 1 \emph{hearts}, 1 \emph{spades} are all the same. The probability of bidding 1 \emph{no trump} is even lower. Therefore, the probability that a player can bid 1 \emph{clubs} safely on the basis of their own hand is 
$
\frac{\binom{13}{10}\binom{39}{3}}{\binom{52}{13}} \approx 4.11606\times10^{-6}
$
(Here, we assume one of the worst-case scenarios, the bidder has three cards of three different suits, and the rest of them are clubs.)
This clearly shows that the same player will call consecutively twice on the basis of his own hand will be significantly less, so the opponent, as well as the teammate, is always dynamic.\\
It is interesting to note that the points of \emph{undertrick} are half of the original auction bridge. This is done because of the absence of a fixed teammate.

\section{{Proposed Scoring Scheme}}
\begin{figure*}[!htb]
    \centering
    \includegraphics[width=0.7\textwidth]{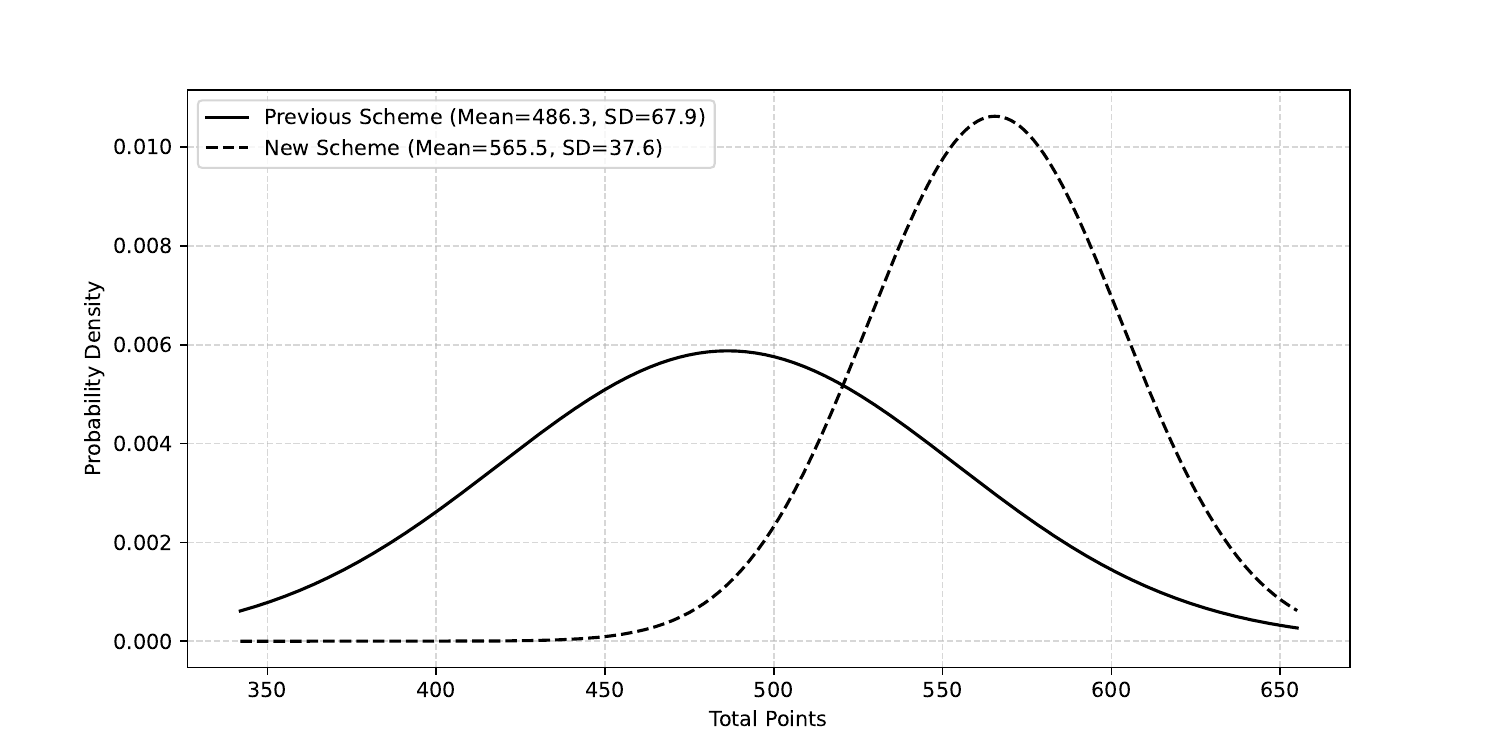}
    \caption{Normal Distribution: Previous vs New Point Scheme}
    \label{Normal Distribution: Previous vs New Point Scheme}
\end{figure*}

We observe that blindly following the pointing scheme \emph{imposed} by the \emph{original} Auction Bridge, the \emph{bidder} gets a very low score. Compared to that, the opponents get a higher score if the \emph{bid or contract} isn't fulfilled. To avoid this \emph{bias}, we slightly tweak the rules to make game slightly more beneficial for the \emph{bidder}. This will also encourage other players to participate in the auction to get a \emph{higher reward} by taking a \emph{higher risk}. In this scheme, no point for the \textbf{bidder} will be halved, as well he will receive \emph{half of the original point} per overtrick. It is done as he doesn't technically have any teammates, as well as it reduces the deviation of points between the players. A consolidated comparative study of previous and new pointing schemes is given in Table~\ref{table_prev_vs_new_point} while Figure~\ref{Normal Distribution: Previous vs New Point Scheme} represents this comparison graphically.
\begin{table*}[!htb]
    \centering
    \caption{Previous Pointing Scheme VS New Pointing Scheme}
    \resizebox{1\textwidth}{!}{ 
    \begin{tabular}{ccccccccccc}
        \toprule
        \textbf{Sl no} & \textbf{Bidder} & \textbf{Trump} & \multicolumn{3}{c}{\textbf{Previous Pointing Scheme}} & \multicolumn{3}{c}{\textbf{New Pointing Scheme}} & \textbf{Win/Loss} & \textbf{Double/redouble} \\
        \cmidrule(lr){4-6} \cmidrule(lr){7-9}
        & & & \textbf{Player A} & \textbf{Player B} & \textbf{Player C} & \textbf{Player A} & \textbf{Player B} & \textbf{Player C} & & \\
        \midrule
        1  & B & 3 Hearts   & 0   & 24  & 0   & 0   & 80  & 0   & Win  & No \\
        2  & C & 1 Hearts   & 0   & 0   & 8   & 0   & 0   & 68  & Win  & No \\
        3  & C & 2 Diamond  & 25  & 25  & 0   & 25  & 25  & 0   & Loss & No \\
        4  & A & 2 Spades   & 18  & 0   & 0   & 76.5& 0   & 0   & Win  & No \\
        5  & C & 3 Clubs    & 50  & 50  & 0   & 50  & 50  & 0   & Loss & No \\
        6  & C & 2 Spades   & 100 & 100 & 0   & 100 & 100 & 0   & Loss & Double \\
        7  & A & 2 Spades   & 0   & 150 & 150 & 0   & 150 & 150 & Loss & Double \\
        8  & C & 1 Spades   & 0   & 0   & 9   & 0   & 0   & 72  & Win  & No \\
        9  & C & 2 Hearts   & 150 & 150 & 0   & 150 & 150 & 0   & Loss & Double \\
        10 & A & 2 Spades   & 0   & 50  & 50  & 0   & 50  & 50  & Loss & Double \\
        11 & B & 2 Hearts   & 100 & 0   & 100 & 100 & 0   & 100 & Loss & No \\
        12 & B & 4 Spades   & 75  & 0   & 75  & 75  & 0   & 75  & Loss & No \\
        \midrule
        \textbf{Total} & & & 518  & 549  & 392  & 576.5 & 605  & 515  & \\
        \midrule
        \multicolumn{3}{c}{Standard Deviation (SD)} & \multicolumn{3}{c}{67.9} & \multicolumn{3}{c}{37.6} & & \\
        \bottomrule
    \end{tabular}
    } \label{table_prev_vs_new_point}
\end{table*}
From Figure~\ref{Normal Distribution: Previous vs New Point Scheme}, it is evident that the result obtained by the \emph{new pointing scheme} is slightly positively skewed, and leptokurtic as well. This suggest that the game is slightly \emph{biased} towards the bidder, offering high reward for taking a high risk. 
\section{Bidding Strategies}
It is interesting to observe that the unique part in this setup is the \emph{bidding} and the adjusted \emph{pointing scheme}. After the bidding is done, the entire game basically reduces to a normal game of \emph{auction bridge}, with a \emph{phantom player} as a teammate of the \emph{bidder}. So, one of the most crucial part of the gameplay is \emph{bidding}, which will be discussed in greater detail.
\subsection{Strategy 1 (Optimistic) - Winning the First 7 Hands}
In this strategy, the winning bidder aims to win the first seven hands independently of their partner's hand. The game starts with an opponent, and for the bidder to secure the first seven hands, they must win the first trick. This requires the bidder to hold all four Aces. We assume that the winning bidder's partner does not have Aces, Kings, Queens, or Jacks. The possible card combinations required to achieve this are given below:
\\
\begin{table}[H]
    \centering
    \begin{tabular}{ccccc}
        \hline
        Combination & Aces & Kings & Queens & Jacks \\ 
        \hline
        1 & 4 & 3 &  &  \\ 
        2 & 4 & 2 & 1 &  \\ 
        3 & 4 & 1 & 1 & 1 \\ 
        \hline
    \end{tabular}
\end{table}
The probability of winning the first seven hands continuously and losing the rest is approximately $14 \times 10^{-5}$
\subsection{Strategy 2 (Optimistic) - Winning the First 8 Hands}
In this strategy, the goal is to win the first 8 hands without depending on the partner’s hand. The possible card combinations required to achieve this are given below:
\begin{table}[H]
    \centering
    \caption{Possible Combinations}
    \begin{tabular}{cccccc}
        \hline
        Combination & Aces & Kings & Queens & Jacks & 10s \\ 
        \hline
        1 & 4 & 4 &  &  &  \\ 
        2 & 4 & 3 & 1 &  &  \\ 
        3 & 4 & 2 & 2 &  &  \\ 
        4 & 4 & 2 & 1 & 1 &  \\ 
        5 & 4 & 1 & 1 & 1 & 1 \\ 
        \hline
    \end{tabular}
\end{table}

The probability of winning the first 8 hands continuously and losing the rest is approximately $3.6 \times 10^{-5}$
{\subsection{Strategy 3 (Optimistic) - Winning the First 9 Hands}}

In this strategy, the goal is to win the first 9 hands without depending on the partner’s hand. The possible card combinations required to achieve this are given below:

\begin{table}[H]
    \centering
    \begin{tabular}{ccccccc}
        \hline
        Combination & Aces & Kings & Queens & Jacks & 10s & 9s \\ 
        \hline
        1 & 4 & 4 & 1 &  &  &  \\ 
        2 & 4 & 3 & 2 &  &  &  \\ 
        3 & 4 & 2 & 2 & 1 &  &  \\ 
        4 & 4 & 2 & 1 & 1 & 1 &  \\ 
        5 & 4 & 1 & 1 & 1 & 1 & 1 \\ 
        \hline
    \end{tabular}
\end{table}

The probability of winning the first 9 hands continuously and losing the rest is approximately $0.5 \times 10^{-5}$ 
\subsection{Missing Suit Advantage}
\textbf{No Trump Call:}
If a suit is missing in any player’s hand, then there is some risk in calling no Trump. If that player wins the call, then automatically their opponent will start the game. If the same suit card is missing or a low card is present in the hand of the bidder's partner and the opponent starts the game using that suit, then the bidder loses back-to-back suits and finally loses the game.

\textbf{Fixed Trump Call:}
If a suit is missing from any player's hand, then the player checks all 13 cards. If any of the suits has more than five cards with some high points, then a fixed trump will be the better option. In the game, a minimum of 1 or 2 rounds will be played using that missing suit card, so they can easily push a lower trump to win the trick. If the bidder’s partner wins the trick, then the bidder tries to push a card of another suit, which also helps the bidder to win more tricks using the same policy.

{ 
\subsection{Importance of Pass in 3-Player Auction Bridge}
}
In 3-Player Auction Bridge, due to the dynamic team structure and different scoring schemes, a player can take some risks based on the fourth unrevealed hand or choose to pass on their call, {  which in turn calls for different types of gaming strategies, as discussed below:}

 \textbf{Defensive strategy:}
All three players start their call based on the 13 cards in their hand, while the fourth hand remains unrevealed. A defensive player always focuses on their own hand. If the bidder is confident in their hand, they may place a bid; otherwise, they may choose to pass.

\textbf{Attack strategy:}
In the attack strategy, the bidder makes a call not only based on their own hand but also considering the fourth unrevealed hand. If the bidder is aware of the calls made by the other two players and sees that their calls depend on their own hands, then only two suits remain. The attacking strategy involves bidding for a suit among these suits with some expectation on the unrevealed hand. The bidder can win the game only if the remaining players do not employ the bluff strategy during the bidding phase. This is also known as a high-risk strategy. It is not independent, but it depends on several factors. Only when all factors align can a bidder win the game without holding a high card in their hand.

\textbf{Bluff strategy}
A bidder using an attacking approach may start bidding with any suit present in their hand, even with very low values or no points at all. If one of the other two bidders has a high card or a significant number (greater than 5 or 6) of cards in the same suit, they may opt for a bluff strategy by passing the call without taking any risk. Before the bidding ends, they may call "double" to maximize their points and win a large score.

{  Based on the above discussion, a game is discussed in Table~\ref{table_2_game_eg}, where \textbf{Player A} uses a \textbf{defensive approach} and \textbf{Player C} uses an \textbf{attacking approach}.}
\begin{table*}[!b]
\caption{Game Data Table}
\label{table_2_game_eg}
\centering
\resizebox{\textwidth}{!}{
\begin{tabular}{c c c c c c c c}
\hline
Sl. No & Bidder & Trump & A Score & B Score & C Score & Outcome & Double \\
\hline
1  & B & 3 Hearts   & 0    & 80   & 0    & Win  & No \\
2  & C & 1 Hearts   & 0    & 0    & 68   & Win  & No \\
3  & C & 2 Diamonds & 25   & 25   & 0    & Loss & No \\
4  & A & 2 Spades   & 76.5 & 0    & 0    & Win  & No \\
5  & C & 3 Clubs    & 50   & 50   & 0    & Loss & No \\
6  & C & 2 Spades   & 100  & 100  & 0    & Loss & Double \\
7  & A & 2 Spades   & 0    & 150  & 150  & Loss & Double \\
8  & C & 1 Spade    & 0    & 0    & 72   & Win  & No \\
9  & C & 2 Hearts   & 150  & 150  & 0    & Loss & Double \\
10 & A & 2 Spades   & 0    & 50   & 50   & Loss & Double \\
11 & B & 2 Hearts   & 100  & 0    & 100  & Loss & No \\
12 & B & 4 Spades   & 75   & 0    & 75   & Loss & No \\
13 & C & 3 Hearts   & 25   & 25   & 0    & Loss & No \\
14 & C & 3 Hearts   & 0    & 0    & 188  & Win  & Double \\
15 & B & 2 Diamonds & 0    & 120  & 0    & Win  & No \\
\hline
\end{tabular}
}
\end{table*}

\section{Exploring Different Game Plans}
The players of the proposed auction bridge game may adopt either of the four game plans: \textbf{\textit{High Card First (HCF)}}, \textbf{\textit{Low Card First (LCF)}}, \textbf{\textit{General Strategy(GS}}), \textbf{\textit{Defeat Seeking Strategy (DSS)}}
The algorithms and and the complexity analysis of these game plans as well as a general strategy are discussed below:
\subsection{High Card First Strategy}
A player using the High Card First strategy always tries to play the highest card first to win consecutive tricks. This strategy is beneficial in the early game, but in the end game, the player may lose many tricks if the opponent is not using the same strategy. The corresponding step by step working rule is given in Algorithm 1.

\begin{algorithm}
\caption{Strategy: High Card First}
\begin{algorithmic}[1]
\State \textbf{Input:} Player's hand, current trick
\State \textbf{Output:} Card to play

\If{trick is empty}
    \State \Comment Player is leading the trick
    \State $selectedCard \gets$ highest-ranking card in hand
    \State \Return $selectedCard$
\Else
    \State \Comment Player is not leading
    \State $leadSuit \gets$ suit of first card in trick
    \State $playableCards \gets$ cards in hand matching $leadSuit$
    
    \If{$playableCards$ is not empty}
        \State $selectedCard \gets$ highest-ranking card in $playableCards$
    \Else
        \State $selectedCard \gets$ lowest-ranking card in hand
    \EndIf
    
    \State \Return $selectedCard$
\EndIf
\end{algorithmic}
\end{algorithm}

Here the "lead suit" is the suit of the first card played in the round.

\subsection{Complexity Analysis: High Card First}
\begin{enumerate}
    \item If the player is the first in the trick:
    \begin{itemize}
        \item Finding the highest-ranked card takes \(O(n)\).
    \end{itemize}
    \item If not the first player:
    \begin{itemize}
        \item Filtering playable cards (matching the lead suit) takes \(O(n)\).
        \item Finding the highest-ranked playable card takes \(O(n)\).
        \item If no playable cards, finding the lowest-ranked card from all cards takes \( O(n) \).
    \end{itemize}
\end{enumerate}
\textbf{Overall Complexity:}  
$
O(n) + O(n) = O(n)
$\\
\textbf{Best-case:} \( O(n) \) (if first player).
\subsection{Low Card First Strategy}

A player using the Low Card First strategy always tries to play the lowest card first to win consecutive tricks at the end game. This strategy is beneficial in the late game, but in the early game, the player may lose many tricks if the opponent is not using the same strategy. {  The corresponding step by step working rule is given in Algorithm 2.}

\begin{algorithm}
\caption{Strategy: Low Card First}
\begin{algorithmic}[2]
\Require Player's hand, current trick
\Ensure Card to be played

\If{trick is empty}
    \State \Comment Player is going first
    \State $\text{cardToPlay} \gets \text{lowest-ranked card in hand}$
    \State \Return $\text{cardToPlay}$
\Else
    \State \Comment Player is not going first
    \State $\text{leadSuit} \gets \text{suit of first played card}$
    \State $\text{matchingCards} \gets \text{cards in hand matching } \text{leadSuit}$
    
    \If{$\text{matchingCards} \neq \emptyset$}
        \State $\text{cardToPlay} \gets \text{lowest-ranked card in } \text{matchingCards}$
    \Else
        \State $\text{cardToPlay} \gets \text{lowest-ranked card in hand}$
    \EndIf
    
    \State \Return $\text{cardToPlay}$
\EndIf
\end{algorithmic}
\end{algorithm}
The overall complexity and best case scenario in this case is also $O(n)$.

\subsection{General Strategy}

A player using the General Strategy always tries to play aces first if available to win a trick. Otherwise, there are two possibilities:

\begin{itemize}
\item \textbf{Case 1:} If no aces are available in the player’s hand and they must start the game, then they should play any low card to rotate the game.
\item \textbf{Case 2:} If no aces are available in the player’s hand and another player starts the game, then they should play the highest card from the same suit to win the trick. If they do not have a card of the same suit, they should play any low card from a different suit.
\end{itemize}

{  The corresponding step by step working rule is given in Algorithm 3.}
\begin{figure*}[!b]
    \centering
    \includegraphics[width=0.75\textwidth]{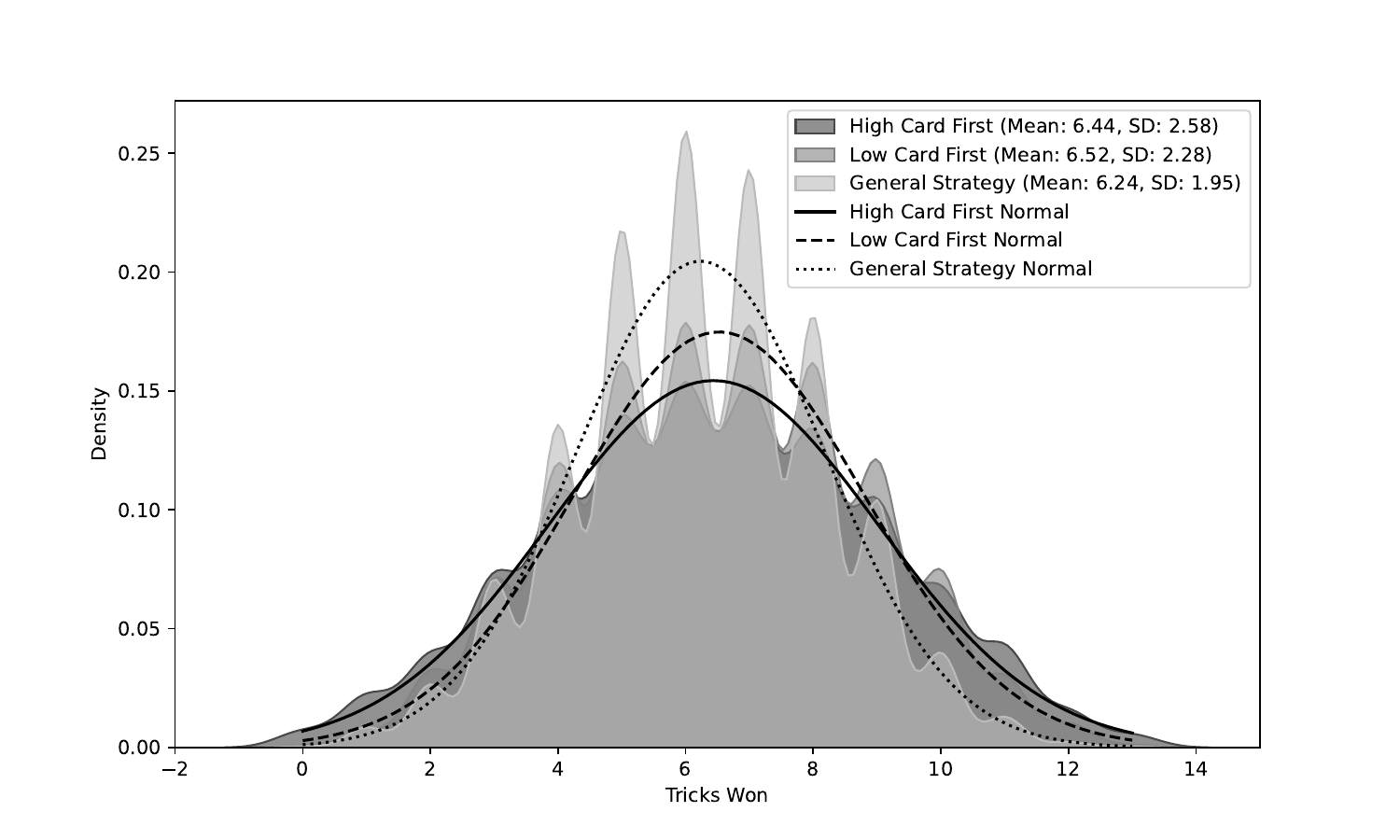} 
    \caption{Distribution of Tricks Won for All Strategies with Normal Fit}
    \label{fig_strategy_compare}
\end{figure*}
\begin{algorithm}
\caption{The General Strategy}
\begin{algorithmic}[1]
\Require Player's hand, current trick
\Ensure Card to be played

\If{trick is empty}
    \State \Comment Leading the trick
    \State $\text{cardToPlay} \gets \text{highest-ranked card in hand}$
    \State \Return $\text{cardToPlay}$
\Else
    \State \Comment Following the trick
    \State $\text{leadSuit} \gets \text{suit of first played card}$
    \State $\text{matchingCards} \gets \{c \in \text{hand} \mid \text{suit}(c) = \text{leadSuit}\}$
    
    \If{$\text{matchingCards} \neq \emptyset$}
        \State $\text{winningCard} \gets \text{highest card in trick so far}$
        \State $\text{beatingCards} \gets \{c \in \text{matchingCards} \mid \text{rank}(c) > \text{rank}(\text{winningCard})\}$
        
        \If{$\text{beatingCards} \neq \emptyset$}
            \State $\text{cardToPlay} \gets \text{highest-ranked card in } \text{beatingCards}$
        \Else
            \State $\text{cardToPlay} \gets \text{lowest-ranked card in } \text{matchingCards}$
        \EndIf
    \Else
        \State $\text{cardToPlay} \gets \text{lowest-ranked card in hand}$
    \EndIf
    
    \State \Return $\text{cardToPlay}$
\EndIf
\end{algorithmic}
\end{algorithm}
A graphical comparison between all the  startegies, with a normal distribution fit, is given in Figure~\ref{fig_strategy_compare}

The overall complexity and best case scenario in this case is also $O(n)$.
Apart from the strategic game plans discussed above, one may consider the following strategies as well.
\section{Some Other Game-play Strategies and the Proposed Defeat Seeking Strategy}
\begin{enumerate}
    \item If a player (Say player A) has significant number of cards of two different suits and the game is NOT a no trump game, then a new approach can be taken. If any other player \emph{bids} one of the suits of player A, then player A would call for the another suit. If the another player gives even higher bid of the first suit, a \emph{double} challenge can be given.
    \item If a player has \emph{King} of a suit and another honors card, and if the player does not have \emph{Ace}, then the another honors card should be played in order to draw the king out and secure the \emph{king}.
    \item If the bidder has significantly less number of card of a particular suit in the \emph{partner} hand, then that suit should be played first.
    \item  If the \emph{bidder} does NOT have a particular suit in hand, and the partner does NOT have another suit in hand, then the \emph{bidder} should alternatively play between two hands, by playing \textit{trump} card from own hand to secure a \emph{trick}, and playing \textit{trump} card from the another hand to secure the other \emph{trick}.
\end{enumerate}
Under the light of all these discussions, it is pretty evident that, in the initial stages, the outcome of the game might be dependent on luck, i.i dependent on the cards you have got as well as the dominant \emph{suit} in your hand. But as the game progresses, one can apply a \emph{optimal} application of several \emph{bidding strategies} and several \emph{gameplay strategies}. Then it mostly becomes a game of \emph{skill}.

\subsection{The Proposed Defeat Seeking Strategies}

In game theory, a defeat-seeking strategy refers to a strategic approach where a player intentionally makes moves that seem to lead to their own defeat but are actually aimed at achieving a long-term advantage {  (refer Straffin~\cite{straffin1993game})}. In this particular context, the \emph{defeat seeking strategy} can be used in many ways. \\
In some cases, if one of the players can deduce any potential information about the hand of any other player, the player might intentionally lose a \emph{trick} to push the gameplay to a particular player in order to get a long term benefit. The another possible strategy can be negative utility exploitation, in which case one of the players deliberately loses for the benefit of another player. As an example, if in a game of \emph{no trump}, player B plays for player A, and player A plays \emph{ace} of a certain suit, then player B will play the highest possible card of the same suit.
This certainly has a similarity with \emph{high card first} strategy. Also, in some cases, if player B has the potential to win a \emph{trick}, player B will still play another card to ensure the victory of player A. This has certain similarities with the \emph{low card first} strategy.\\
But this \emph{defeat seeking phenomena} becomes pretty important when the game revolves around trump cards. If player B has the potential to win a trick by playing a \textit{trump card} which otherwise would have been won by player A, player B would still play another \emph{non-trump card} to let player A win. Then when the game revolves around trump cards only, the defeat-seeking player (i.e. player B) shifts to the strategy mentioned in the \emph{no-trump case}. This assures that player A wins the game by a significant margin.

\subsection{Defeat Seeking Strategy: Static vs Dynamic Team}

In the 4-player game, fixed partners and static teams are available, so a defeat-seeking strategy is not applicable if and only if the fixed partner wants to lose the game.

In the 3-player game, the team and the partner are both dynamic. Thus, a defeat-seeking strategy plays a significant role in that game.

\subsection{Role of Defeat-Seeking Player}

If a defeat-seeking player always wants a specific player to win except himself, then that player always tries to win the call because he has started the game with the advantage of 3 hands. If he loses the call, then the defeat seeker player changes his strategy; otherwise, another player who won the call wins the game.

For example, consider Player C playing as a defeat-seeking strategy and Player A taking support from Player C. If Player B wins the call, both A and C suffer. If Player B loses the match, then Player A and C both win some points. However, as a defeat-seeking player, Player C does not need to win points; he only wants Player A to win more and more points. Consequently, Player A always tries to win the call with a high probability of winning the game. If Player A loses the call, then the defeat seeker player changes his strategy from defeat-seeking to a general strategy so that Player B does not have a 3-hand advantage.

\subsection{No Trump and Fixed Trump}

In a no-trump call or fixed-trump call, defeat-seeking players always try to play high-card first to help their unofficial partner if that player wins the call. If the player loses the call, the defeat-seeking player again shifts from defeat-seeking to a general strategy. Therefore, in a no-trump call, defeat-seeking is quite similar to a high-card first or general strategy.

For example, before starting the game, suppose Player A wins the call with 2 hearts, and Player C plays the defeat-seeking role and supports A. However, according to game rules, Player B and Player C play as partners, making Player A the unofficial partner of Player C. If Player A plays an Ace of Hearts and Player C has a King of Hearts along with other hearts, as a defeat-seeking player, he should play the King of Hearts instead of any other heart. This decision confirms that the King of Hearts trick is lost and provides Player A with a 1-trick advantage.

A \textbf{Fixed Trump} strategy always provides an advantage for the bidder. However, the bidder does not know their partner's hand. There are three possible types of hands the partner might have:

\begin{enumerate}
    \item \textbf{Good Hand}: A good number of trump cards and a good number of face cards.
    \item \textbf{Supportive Hand}: Either a smaller number of trump cards with a good number of face cards, or a good number of trump cards with a smaller number of face cards.
    \item \textbf{Bad Hand}: A smaller number of both trump and face cards.
\end{enumerate}

If opponents are not engaging in bluffing, and their calls are based on suit colors (i.e., playing a \textbf{No Trump} strategy), then the bidder has two remaining suits to choose from. In this case, the bidder will select the suit that:
\begin{itemize}
    \item contains high cards,
    \item And has the maximum number of cards present in their hand.
\end{itemize}

Now, the presence of the same suit cards with high ranks in the partner’s hand depends on \textbf{conditional probability}.
A \textbf{Fixed Trump} strategy is always a better approach than a \textbf{No Trump} strategy in 3-player Auction Bridge. If the bidder’s partner has either a supportive or good hand, then the probability of winning the match is high (though it depends on several factors). However, in a \textbf{No Trump} game, the partner’s hand must be good, otherwise the bidder is likely to lose the game.
This represents one of the key differences between \textbf{4-player} and \textbf{3-player} Auction Bridge.

\subsection{Equivalence to the Ruin Problem}
Let us say in a particular game, a particular player has scored significantly high than the other two players. Now, if the player has no \emph{honors card} of a particular suit, but other cards of the same suit are at a significant number, then the player might take a \emph{risk} to give a higher call. If he wins, he is benefited, otherwise, the rest of the players gain some point, but not enough to \emph{overpower} the one with highest point. The player with the highest point can do the same while challenging the other player for a \emph{double} or \emph{redouble}. This can be thought of as a $n$-step profitable game, where the player with the highest point has enough \emph{money} to spare to try his or her luck, but the players with significantly lower points can't usually take such risks.

\section{Some Examples}
\afterpage{
\begin{table*}
\caption{Final Results of Strategies and Team Scores}
\label{table:example_4}
\centering
\setlength{\tabcolsep}{3pt}
\resizebox{1.0\textwidth}{!}{
\begin{tabular}{c c c c c c c}
\hline
\textbf{Strategy} & \textbf{Player 0} & \textbf{Player 1} & \textbf{Player 2} & \textbf{Player 3} & \textbf{Team A (0\&2)} & \textbf{Team B (1\&3)} \\
\hline
High Card First    & 0 & 4 & 6 & 3 & 6 & 7 \\
Low Card First     & 1 & 4 & 5 & 3 & 6 & 7 \\
General Strategy   & 3 & 2 & 4 & 4 & 7 & 6 \\
\hline
\end{tabular}%
}
\end{table*}
}
\subsection{Example 1: Random Game with 3 Different Strategies}

\begin{itemize}
    \item\textbf{{Player 0:}}
        (6, S), (8, S), (Q, S)
        (2, H), (3, H), (Q, H)
        (6, D), (Q, D)
         (3, C), (4, C), (5, C), (7, C), (J, C)
    \item\textbf{{Player 1:}}
     (5, S), (7, S), (K, S)
     (4, H), (10, H), (A, H)
     (4, D), (5, D), (8, D), (K, D)
     (9, C), (10, C), (A, C)

    \item\textbf{{Player 2:}}
    (2, S), (3, S), (4, S)
     (5, H), (9, H), (J, H), (K, H)
     (7, D), (10, D), (J, D), (A, D)
     (6, C), (Q, C)
    \item\textbf{{Player 3:}}
    (9, S), (10, S), (J, S), (A, S)
     (6, H), (7, H), (8, H)
     (2, D), (3, D), (9, D)
     (2, C), (8, C), (K, C)
\end{itemize}
{Here, (6,S) stands for 6 of Spades, and the rest follows in similar fashion. The final results from the above game are given in Table~\ref{table:example_4}}.
\subsection{A Simulated Game}
Before starting the game, in the calling period, all the 3 players can bid 1,2,3,\ldots, 7 No Trump calls (Right now we are only focusing on No Trump call). The winning bidder should collect (The number he called +6) tricks with his partner into winning the game.

\subsection{Most Likely Strategy to Bid 1, 2, and 3 No Trump}
Let us assume Aces carry 5 points, King carry 4 points, Queens carry 3 points, Jacks carry 2 points, 10's carry 1 point, Rest of all cards carries 0 point.

\textbf{Strategy 1 for calling:} 
A player calls 1 No Trump with 20 or more points, 2 No Trump with 25 or more points, and 3 No Trump with 30 or more points.

\textbf{Strategy 2 for calling:} 
In this approach, a player calls 1 No Trump only with 25 or more points, 2 No Trump with 30 or more points, and 3 No Trump with 35 or more points.

Let us assume, based on both strategies Player 2 always wins the call among 1, 2, 3, and automatically off-hand will play as Player 2’s partner. Player 1 and 3 play as team A, and player 2 and off-hand play as team B.

Now after 1,000,000 simulations, the result is shown in a matrix format. For both cases, 3 times simulation data will be presented below.

\subsection{Simulation Results}

\begin{table}[H]
    \centering
    \caption{Comparison of No Trump Bidding Rules}
    \begin{tabular}{ccccc}
    \toprule
    Rule Set & Simulation & 1NT & 2NT & 3NT \\
    \midrule
    \textbf{20+/25+/30+ pts} & \textbf{Simulation 1} & & & \\
    & Calls & 157499 & 38669 & 4858 \\
    & Yes   & 117599 & 26844 & 3234 \\
    & No    &  39900 & 11825 & 1624 \\
    \cmidrule{2-5}
    & \textbf{Simulation 2} & & & \\
    & Calls & 157548 & 38736 & 4833 \\
    & Yes   & 117750 & 27104 & 3176 \\
    & No    &  39798 & 11632 & 1657 \\
    \cmidrule{2-5}
    & \textbf{Simulation 3} & & & \\
    & Calls & 157535 & 38763 & 4941 \\
    & Yes   & 117682 & 27223 & 3284 \\
    & No    &  39853 & 11540 & 1657 \\
    \midrule
    \textbf{25+/30+/35+ pts} & \textbf{Simulation 1} & & & \\
    & Calls & 38365 & 4528  & 268 \\
    & Yes   & 33326 & 3876  & 227 \\
    & No    &  5039 &  652  & 41  \\
    \cmidrule{2-5}
    & \textbf{Simulation 2} & & & \\
    & Calls & 38631 & 4641  & 265 \\
    & Yes   & 33578 & 3958  & 210 \\
    & No    &  5053 &  683  & 55  \\
    \cmidrule{2-5}
    & \textbf{Simulation 3} & & & \\
    & Calls & 38409 & 4587  & 265 \\
    & Yes   & 33492 & 3834  & 222 \\
    & No    &  4917 &  753  & 43  \\
    \bottomrule
    \end{tabular}
\end{table}
\section{Dynamic Programming vs Enumerative Algorithm}

\subsection{Dynamic Programming}
Using Dynamic Programming, a player tries to remember all the cards that have already been played. Based on that, the player will play a card from his hand to win that trick or skip the trick to win future tricks. Backtracking is the primary technique we can use in the Auction Bridge Game.

\subsection{Enumerative Algorithm}
Using the Enumerative algorithm, we can easily solve the problem. If a player starts the trick, he should play a high-point card to win the trick. If someone else starts the game, then that specific player checks his hand. If a higher card exists, he should play that one; otherwise, he will play the lowest card from the same suit if the same suit card is present in his hand.

\subsection{Randomly 10 Simulation Data}
Randomly choose 13 cards out of 52 for player 2, then choose 13 cards from 39 for player 4. Now, the remaining 26 cards should be distributed between players 1 and 3 in $\binom{26}{13}$ = 10400600 ways. Players 1 and 3 mean Team A, and Players 2 and 4 mean Team B. Now, Team B can win tricks 0, 1, 2, ..., 13. We calculate the frequency of all tricks in the Table~\ref{table:simulated_hands}.

\begin{table*}
\caption{Simulated Hands for Player 2 and Player 4}
\label{table:simulated_hands}
\centering
\renewcommand{\arraystretch}{1.2}
\begin{tabular}{c p{7.2cm} p{7.2cm}} 
\hline
\textbf{Simulation No} & \textbf{Player 2 Hand} & \textbf{Player 4 Hand} \\
\hline
1 & [2H, 2S, 5D, 5H, 6D, 7H, 8C, 8S, 9C, JC, QC, TH, TS] & [2D, 3C, 3D, 3H, 4D, 6H, 7D, 7S, AH, AS, JS, KD, QD] \\
2 & [4D, 4S, 6D, 6H, 7S, 8C, 8D, 9H, AD, JD, QC, TH, TS] & [3H, 4C, 5S, 6C, 6S, 8H, 9D, AC, JH, JS, KC, QS, TC] \\
3 & [2D, 2S, 3D, 3H, 4C, 5H, 6S, 7S, 8D, 9C, AH, JD, TD] & [2C, 4S, 5C, 5S, 7D, 8S, AD, AS, JH, JS, KC, KS, QH] \\
4 & [2H, 2S, 3H, 4S, 5H, 7C, 7H, 8C, AC, JD, JH, QH, TH] & [4D, 4H, 5C, 6D, 8D, AH, JC, KH, KS, QD, QS, TC, TS] \\
5 & [2H, 2S, 4D, 7C, 7D, 8H, 9S, AH, JS, KC, KS, QD, TH] & [2D, 3D, 4C, 4H, 5C, 6C, 9C, 9D, JD, QC, QS, TC, TD] \\
6 & [2D, 2S, 3D, 4C, 6D, 8H, 8S, 9C, AS, QD, QH, TC, TS] & [3C, 3H, 4D, 5C, 6H, 7H, AC, JD, JS, KC, QS, TD, TH] \\
7 & [2H, 3D, 5D, 5S, 8H, 8S, 9S, JC, JS, KC, QC, TC, TS] & [2D, 2S, 3C, 4C, 5C, 7D, 7H, 7S, 8D, 9D, AC, KH, TH] \\
8 & [2C, 3H, 3S, 6S, 7D, 9C, 9S, AC, JD, KC, QD, TH, TS] & [2D, 2H, 5C, 5D, 6C, 6D, 7S, AS, JC, JS, QC, TC, TD] \\
9 & [2D, 2S, 3D, 4D, 4H, 5C, 5D, 8D, JS, KC, KD, TD, TS] & [2C, 3C, 3H, 6H, 6S, 8S, 9C, 9D, JH, KH, QC, QS, TH] \\
10 & [2S, 3S, 4C, 4S, 6C, 6H, 7D, 8H, 9S, AS, JC, JH, TH] & [2C, 4D, 4H, 5D, 5S, 6D, 6S, 7S, 8C, 8D, AC, KC, TC] \\
\hline
\end{tabular}
\end{table*}

\section{Future Scopes and Possible Extensions}
To further analyze the proposed game and its strategic implications, concepts from game theory involving three-player interactions might be introduced. Unlike traditional two-player zero-sum games, this variant presents a more complex scenario where each player’s decision influences both their own outcome and the outcomes of multiple opponents. Understanding the equilibrium strategies and optimal decision-making in such a setting could provide valuable insights into the competitive dynamics of the game. Studying Nash equilibria in this multi-player environment may reveal whether players have dominant strategies or if mixed strategies are required for optimal play.  

A key aspect of this analysis would involve determining whether the proposed game is a zero-sum game or not. In classical game theory, a zero-sum game is one where the total gain and loss across all players always sum to zero, meaning one player’s gain comes directly at the expense of others. However, in this modified bridge variant, the reward distribution and the adjustments made to reduce bias might lead to a different classification. If the game is not zero-sum, it would imply that cooperative or semi-cooperative strategies might emerge, which would significantly alter the dynamics of bidding and gameplay. A formal proof or counterexample would be crucial in settling this question, adding theoretical depth to the study.  

Additionally, structured training procedures could be explored to develop and refine strategies for optimal gameplay. AI models or human players could be trained using different approaches, such as a balanced general strategy aimed at maximizing expected value or an aggressive defeat-seeking strategy that prioritizes minimizing opponents' gains. Machine learning techniques, such as reinforcement learning, could be applied to simulate thousands of games and optimize decision-making over time. Understanding these strategies could not only improve human gameplay but also provide a framework for AI-driven competitive play.


\bibliographystyle{plain}
\bibliography{refs}
\end{document}